\documentclass[preprint,12pt]{elsarticle}

\usepackage{graphicx}
\usepackage{amssymb}

\journal{Nuclear Physics A }

\begin{document}

\begin{frontmatter}

\title{Nuclear medium modification of the $F_{2}(x, Q^2)$ structure function}

\author{M. Sajjad Athar \corref{cor1}}
\ead{sajathar@gmail.com}
\cortext[cor1]{Corresponding author}
\address{Department of Physics, Aligarh Muslim University, Aligarh-202 002, India}

\author{I Ruiz Sim\'o and M J Vicente Vacas}

\address{Departamento de F\'{\i}sica Te\'orica and IFIC, \\
Centro Mixto Universidad de
Valencia-CSIC,\\
46100 Burjassot (Valencia), Spain}

\begin{abstract}
We study the nuclear effects in the electromagnetic structure function  $F_{2}(x, Q^2)$  in the deep inelastic lepton nucleus scattering process by taking into account Fermi motion, binding, pion and rho meson cloud contributions. Calculations have been done in a local density approximation using 
relativistic nuclear spectral functions which include nucleon correlations. The ratios $R_{F2}^A(x,Q^2)=\frac{2F_2^A(x,Q^2)}{AF_2^D(x,Q^2)}$ are obtained and compared with  recent JLab results for light nuclei  with special attention to the slope of the $x$ distributions. This magnitude  shows a non trivial A dependence and it is  insensitive to possible normalization uncertainties. 
The results  have also been compared with some of the older experiments using  intermediate mass nuclei.

\end{abstract}

\begin{keyword}
Structure function \sep Nuclear medium effects \sep Deep inelastic scattering \sep Local density approximation 



\end{keyword}

\end{frontmatter}


\section{Introduction}
Recently Jefferson Lab(JLab)~\cite{Seely} using a high intensity electron beam of energy 5.767 GeV has measured the nuclear dependence of the structure function in some nuclei by 
studying the ratio R($x,Q^2$)=$\frac{2\sigma^A}{A\sigma^D}$, where $\sigma^A$ is the inclusive cross section in nuclei and $\sigma^D$ is the inclusive cross section in deuterium. 
The experimental results for the ratio R($x,Q^2$) have been presented by them~\cite{Seely} for 0.3 $< x <$ 0.9 and have re-confirmed the older EMC results~\cite{Aubert:EMC,Arnold:EMC,Bodek:EMC} 
that the structure function of a nucleon is modified when it is placed inside a nucleus.  This experiment shows that the slope of the EMC effect does not scale with the nuclear density and therefore the simple models to implement these nuclear effects, based on A or 
average density fits, for example as described by Gomez et al.~\cite{Gomez:SLAC}, fail to describe the new and precise results for light nuclei.

The behavior of $R_{F2}^A(x,Q^2)$ can be broadly divided into four categories viz. $x \le 0.1$ is the shadowing region, 0.1$\le x\le$0.3 is the anti-shadowing region, 0.3$\le x\le$0.8 is the 
EMC region and beyond $x\approx 0.8$, known as the Fermi motion region. Theoretically, many analysis have been done to study the EMC effect and various models have been proposed and discussed 
in the literature~\cite{Arneodo:1994,Geesaman:1995,Armesto:2006ph,Miller:EMCTh}.  Several phenomenological parameterizations for the nuclear parton distribution functions(NPDFs) have been discussed in 
the literature like the works of Hirai et al.~\cite{Hirai:NPDF},
Eskola et al.~\cite{Eskola:NPDF}, Schienbein et al.~\cite{Schienbein:NPDF,Schienbein:2009kk}  which successfully reproduce the nuclear modifications in the deep inelastic lepton-nucleus and neutrino-nucleus 
scattering.

In this work, we study the nuclear medium effects on the structure function within a model based on the theoretical calculation of Ref.~\cite{Marco} with the aim of comparing it with the recent JLab data.
The spectral function that describes the energy and momentum distribution of the nucleons in nuclei is obtained by using the Lehmann's representation for the relativistic nucleon propagator and nuclear 
many body theory is used to calculate it for an interacting Fermi sea in nuclear matter~\cite{FernandezdeCordoba:1991wf}. A local density approximation is then applied to translate these results into finite nuclei. The contributions of the pion and rho meson clouds  are taken into account in a many body field theoretical approach which is  based on Refs.~\cite{Marco,GarciaRecio:1994cn}. 
The model from Ref.~\cite{Marco} has been  improved  in several ways. The old model used the Bjorken limit and assumed the Callan-Gross relationship for nuclear structure functions ${F_2}^A(x)$ and
 ${F_1}^A(x)$. Due to the fact that JLab data have been taken in a region of relatively low $ Q^2 $  
($ Q^2 \sim 3-6 $  GeV$^2$) we have not assumed the Bjorken limit. Also, for low $Q^2$ and moderate 
$x$ values Target Mass Corrections (TMC) might play an important role. We have incorporated them following Ref.~\cite{Schienbein:2007gr}. Another difference with respect to Ref.~\cite{Marco} is the 
fact that for the ratios we  divide  by the deuteron structure function, rather than the nucleon one.
This only implies substantial changes at moderate and high $x$ values.  We have also considered  shadowing because it reduces the contribution coming from the pion and rho meson clouds~\cite{Petti,Petti2}.
For the numerical calculations, next to leading order (NLO) Parton Distribution Functions (PDF) for the nucleons have been taken from the parameterization of Martin et al. (MSTW)~\cite{MSTW}.  
The NLO evolution of the deep inelastic structure functions has been taken from the works of Vermaseren et al.~\cite{Vermaseren} and van Neerven and Vogt~\cite{Neerven}.  In the case of pions 
we have taken the pionic parton distribution functions given by Gluck et al.~\cite{Gluck:1991ey,Gluck}.  For the rho mesons, we have applied the same PDFs as for the pions as in Ref.~\cite{Marco}.

The structure of the paper is as follows: In Sect.~\ref{DIS_lepton_nucleon} we introduce some basic formalism for lepton-nucleon scattering, in Sect.~\ref{sec:NE} we analyse the different nuclear effects,
in Sect.~\ref{sec:DE} we consider the deuteron case and we end by comparing our results with data
in Sect.~\ref{sec:RE}.
\section{Deep inelastic lepton-nucleon scattering } \label{DIS_lepton_nucleon}

\begin{figure}
\begin{center}
\includegraphics{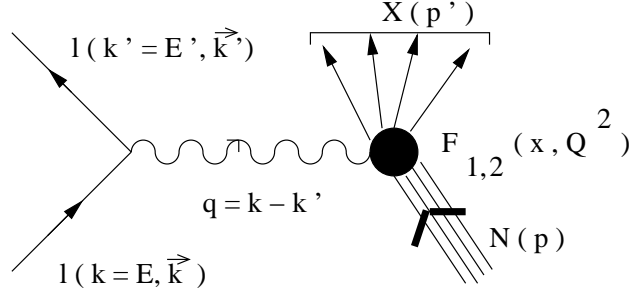}
\caption{Feynman diagram for the deep inelastic lepton-nucleon scattering}
\label{fg:fig1}
\end{center}
\end{figure}

The double differential cross section for the reaction of scattering of a charged lepton from an unpolarized nucleon in the one photon exchange approximation,
\begin{equation} 	\label{reaction}
l^-(k) + N(p) \rightarrow l^-(k^\prime) + X(p^\prime),~l=~e,~\mu
\end{equation}
depicted in Fig.\ref{fg:fig1} is given, in terms of the Bjorken variables $x$ and $y$, by

\begin{eqnarray}\label{diff_dxdy}
\frac{d^2 \sigma}{d x d y}&=&
\frac{8 M E \pi \alpha^2 }{Q^4}
\left\{xy^2 F_1(x, Q^2)
+ \left(1-y-\frac{xyM}{2 E}\right) F_2(x, Q^2)
\right\}\,,
\end{eqnarray}
where 
\begin{eqnarray}	\label{Bj_var}
x=\frac{Q^2}{2M\nu}, \quad y=\frac{\nu}{E_{l}}
\end{eqnarray}
and $\nu$ is the energy transferred to the hadronic system.
$F_i(x,Q^2)$ are dimensionless structure functions.
In the Bjorken limit, i.e. $Q^2 \rightarrow \infty$, $\nu \rightarrow \infty$, $x$ finite, the structure functions $F_i(x, Q^2)$  depend only on the  variable $x$  and satisfy the Callan-Gross relation~\cite{Callan} given by $2xF_1(x)=F_2(x)$. Using this, the cross section of Eq.(\ref{diff_dxdy}) can be expressed in terms of  $F_2(x)$ and thus the ratio of cross sections is equal to the ratio of structure functions $F_2$.
Even far from the Bjorken limit or when one goes beyond the lowest order (LO), where the Callan-Gross  relation does not hold, the ratio of cross sections  still equals the ratio of structure functions $F_2$ if the ratio of longitudinal to transverse cross sections $R=\sigma_L/\sigma_T$ does not depend on $A$. There is a considerable amount of experimental evidence supporting this fact (e.g. Fig.~6 of Ref.~\cite{Geesaman:1995}). Therefore, in the following we only consider $F_2$ and compare directly $F_2$ ratios with cross section ones.

The nucleon structure functions are determined in terms of parton distribution functions for quarks and anti-quarks.
In this work, for the nucleons we work at NLO~\footnote{On the other hand, the leading order (LO) pionic parton distribution functions of Gluck et al.~\cite{Gluck:1991ey,Gluck} have been used for pions as well as for rho mesons.} and we have used the Parton 
Distribution Functions (PDF) of Martin et al. (MSTW)~\cite{MSTW}. 
At this order, the expression for  the $F_2$ and $F_L$ structure functions can be expressed as  functions of the PDFs by ~\cite{Vermaseren,Neerven,Moch:2004xu}
\begin{equation}\label{F_L}
x^{-1}F_{2,L}=\sum_{f=q,g} C_{2,L} \otimes f, 
\end{equation}
where $C_{2,L}$ are the coefficient functions for the quarks and gluons~\cite{Vermaseren,Neerven,Moch:2004xu} and $f$ represents the quark and gluon distributions~\cite{MSTW}.

\section{Nuclear effects}\label{sec:NE}
We have used the local density approximation (LDA) to incorporate nuclear medium effects\footnote{The nuclear densities have been taken from Ref.~\cite{De Jager:1987qc}.}. Inside the nucleus, when the reaction given by Eq.(\ref{reaction}) 
takes place, several nuclear effects  like Fermi motion, binding, pion and rho meson cloud contributions must be taken into account. Fermi motion and nucleon binding are 
implemented through the use of  a nucleon spectral function. The relativistic nucleon propagator in a nuclear medium can be cast as~\cite{Marco, FernandezdeCordoba:1991wf}:
\begin{eqnarray}\label{Gp}
G (p) = \frac{M}{E({\bf p})} 
\sum_r u_r ({\bf p}) \bar{u}_r({\bf p})
\left[\int^{\mu}_{- \infty} d \, \omega 
\frac{S_h (\omega, \mathbf{p})}{p^0 - \omega - i \eta}
+ \int^{\infty}_{\mu} d \, \omega 
\frac{S_p (\omega, \mathbf{p})}{p^0 - \omega + i \eta}\right]\,,
\end{eqnarray}
 where $S_h (\omega, \mathbf{p})$ and $S_p (\omega, \mathbf{p})$ are the hole
and particle spectral functions respectively. Full details can be found in Ref.~\cite{FernandezdeCordoba:1991wf}. We ensure that the spectral function is properly normalized and we get the 
correct Baryon number for the nucleus. Furthermore, we have also calculated the kinetic energy and the binding energy per nucleon and have found that the theoretical binding energy is very close to the experimentally observed  ones for $^9$Be, $^{12}$C, $^{40}$Ca and $^{56}$Fe.

Our base equation for the nuclear structure function $F_2^A$ in an isoscalar target is:
\begin{eqnarray}\label{F2A_Kulagin}
F^A_2(x,Q^2)&=&4\int d^3r\int \frac{d^3p}{(2\pi)^3}\int_{-\infty}^\mu d\omega\;S_h(\omega,\mathbf{p},\rho(\mathbf{r}))\frac{\left(1-\gamma\frac{p_z}{M}\right)}{\gamma^2}\\ \nonumber
&& \times \left(\gamma'^2+\frac{6x'^2(\mathbf{p}^2-p^2_z)}{Q^2}\right)F_2^N(x',Q^2)
\end{eqnarray}
with $p^0=M+\omega$, $\gamma'^2=1+4x'^2p^2/Q^2$ and $x'$ is $Q^2/(2p\cdot q)$.
This expression is equivalent to that of Ref.~\cite{Petti} after trivial algebraic transformations and taking into account the different normalization of the spectral function $\mathcal{P}_0(\epsilon,\mathbf{p})$ used in 
\cite{Petti} such that
\begin{equation}\label{Relation_between_SpectralFunctions}
A\;\mathcal{P}_0(\epsilon,\mathbf{p})\longrightarrow4\cdot2\pi\int d^3r\;S_h(\omega,\mathbf{p},\rho(\mathbf{r}))\,.
\end{equation}

In an earlier study, the behaviour of different nucleon spectral functions has been analysed~\cite{Simo}. In particular, the  spectral functions given by Fern\'andez de C\'ordoba and Oset \cite{FernandezdeCordoba:1991wf}, Kulagin and Petti \cite{Petti}, and Ankowski et al. \cite{Ankowski:2007uy} were used and compared. It was found that  the results do not change appreciably. Finally, we should comment that  the present formalism has also been used to study the nuclear effects in the $F_3$ structure function~\cite{Sajjad}.

\subsection{$\pi$ and $\rho$ mesons contribution to the nuclear structure function}\label{Pion_Contribution}
\label{sec:meson}

The pion and rho meson cloud contributions to the $F_2$ structure function have been implemented following the many body field theoretical approach of Refs.~\cite{Marco,GarciaRecio:1994cn}. The pion structure function $F_{2 A, \pi} (x)$ is written as

\begin{equation}  \label{F2pion}
F_{2, \pi}^A (x) = - 6 \int  d^3 r  \int  \frac{d^4 p}{(2 \pi)^4} \; 
\theta (p^0) \; \delta I m D (p) \; 
\; \frac{x}{x_\pi} \; 2 M \; F_{2 \pi} (x_\pi) \; \theta (x_\pi - x) \; 
\theta (1 - x_\pi) 
\end{equation}
where $D (p)$ the pion propagator in the medium given in terms of the pion self energy $\Pi_{\pi}$:
\begin{equation}
D (p) = [ p^{0 2} - \vec{p}\,^{2} - m^2_{\pi} - \Pi_{\pi} (p^0, p) ]^{- 1}\,,
\end{equation}
where
\begin{equation}\label{pionSelfenergy}
\Pi_\pi=\frac{f^2/m_\pi^2 F^2(p)\vec{p}\,^{2}\Pi^*}{1-f^2/m_\pi^2 V'_L\Pi^*}\,.
\end{equation}
Here, $F(p)=(\Lambda^2-m_\pi^2)/(\Lambda^2+\vec{p}\,^{2})$ is the $\pi NN$ form factor and  $\Lambda$=1 GeV, $f=1.01$, $V'_L$ is
the longitudinal part of the spin-isospin interaction and $\Pi^*$ is the irreducible pion self energy that contains the contribution of particle - hole and delta - hole excitations. In Eq.(\ref{F2pion}), $\delta Im D(p)$ is given by  

\begin{equation}
\delta I m D (p) \equiv I m D (p) - \rho \;
\frac{\partial Im D (p)}{\partial \rho} \left|_{\rho = 0} \right.
\end{equation}
and
\begin{equation} 
\frac{x}{x_{\pi}} = \frac{- p^0 + p^z}{M}
\end{equation}

Assuming SU(3) symmetry and following the same notation as in Ref.\cite{Gluck:1991ey},  the pion structure function at LO can be written in terms of pionic PDFs as
\begin{equation}\label{pion_structure_function}
F_{2\pi}(x_\pi)=x_\pi\left(\frac{5}{9}\;v_\pi(x_\pi)+\frac{12}{9}\;\bar{q}_\pi(x_\pi)\right)
\end{equation}
where $v_\pi(x_\pi)$ is the valence distribution and $\bar{q}_\pi(x_\pi)$ is the light SU(3)-symmetric sea distribution.

Similarly, the contribution of the $\rho$-meson cloud to the structure function  is written as~\cite{Marco}

\begin{equation} \label{F2rho}
F_{2, \rho}^A (x) = - 12 \int d^3 r \int \frac{d^4 p}{(2 \pi)^4}
\theta (p^0) \delta Im D_{\rho} (p) \frac{x}{x_{\rho}} \, 2 M
F_{2 \rho} (x_{\rho}) \theta (x_{\rho} - x) \theta (1 - x_{\rho})
\end{equation}

\noindent
where $D_{\rho} (p)$ is the $\rho$-meson propagator and $F_{2 \rho}
(x_{\rho})$ is the $\rho$-meson structure function, which we have taken equal to the pion structure function $F_{2\pi}$ using the valence and sea pionic PDFs from reference \cite{Gluck:1991ey}. 
$\Lambda_\rho$ in $\rho NN$ form factor $F(p)=(\Lambda_\rho^2-m_\rho^2)/(\Lambda_\rho^2+\vec{p}\,^{2})$ has also 
been taken as 1 GeV.

Further details concerning the pion and $\rho$-meson propagator can be found in Ref.~\cite{Marco}.
This model for the pion and $\rho$ selfenergies has been abundantly used in the intermediate energy region and provides a quite solid description of a wide range of phenomenology  in pion, electron and photon induced reactions in nuclei, see e.g. Refs. \cite{Oset:1981ih,Carrasco:1989vq,Nieves:1991ye,Nieves:1993ev,Gil:1997bm} and
references in~\cite{Marco}. In particular, a careful study of the in medium pion propagator used here was carried out  in Ref.~\cite{GarciaRecio:1994cn}.
There, several tests concerning the fulfillment of sum rules, and the preservation of the analytical properties of the meson propagator and the consistency of the results with similar calculations  were considered.

In addition, the  balance of light-cone momentum between bound nucleons and pions can be studied by means of a momentum sum rule as done in Ref.~\cite{Petti}. The pion $<y>_\pi$ and nucleon $<y>_N$ fractions of the light cone momentum are related by
\begin{equation}
 <y>_\pi+ <y>_N = \frac{M_A}{A M},
\end{equation}
where $M_A$ is the nucleus mass. See section 5.3 of Ref.~\cite{Petti} for details. The  sum rule should be valid for a nuclear model where the Hamiltonian would contain only pions and nucleons. In fact, our model for the nucleon spectral function is based on a phenomenological approach that also contains many other pieces in the nucleon-nucleon interaction and thus the sum rule is not directly applicable. Nonetheless, it can provide further constraints on the size of the mesonic contribution and it will be discussed in the results section.

The  mesonic cloud  contribution is expected to be negligible for  deuteron as it depends, roughly 
speaking, quadratically on the baryon density which is quite small for this case\footnote{ A direct application of our model to deuteron, produces a mesonic contribution that is always lower than a 0.6 percent of the nucleonic contribution for the analysed $x$ range. Thus, its inclusion would have a very minor effect in the ratios. Nonetheless, we should mention that
our formalism, which  starts from selfenergies calculated in nuclear matter, is not expected to be very reliable for the calculation of the mesonic effects in deuteron.}. Therefore,
these contributions have not been included in the evaluation of the deuteron structure function.

\subsection{Target mass corrections}
Target mass corrections have been incorporated by means of the approximate formula~\cite{Schienbein:2007gr}
\begin{equation}\label{f2TMC}
 F_{2}^{TMC}(x,Q^2)\simeq\frac{x^2}{\xi^2\,\gamma^3} F_{2}(\xi,Q^2)\left[ 1+\frac{6\, \mu\, x\, \xi}{\gamma}(1-\xi)^2\right],
\end{equation}
 where $\mu=\frac{M^2}{Q^2}$, $ \gamma = \sqrt{ 1 + \frac{4 x^2 M^2}{ Q^2 } }$ and $\xi$ is the Natchmann variable defined as
\begin{equation}\label{Natchmann}
\xi = \frac{2 x}{1+\gamma}\,.
\end{equation}

\subsection{Coherent nuclear effects}\label{shadowing}

Furthermore, we have taken into account the shadowing effect following the works of Kulagin and Petti~\cite{Petti}. We are interested in the relative effect in $F_2^A$ that can be written as 
\begin{equation}\label{deltaR2}
\delta R_2=\frac{\delta F^A_2}{F_2^N}=\frac{1+R^2}{1+R}\delta R_T
\end{equation}
where $R(x,Q^2)$ is calculated for the free nucleon. 
For $\delta R_T$ we use expression 63 in Ref.~\cite{Petti}.
The most general expression for $R(x,Q^2)$, taking into account the target mass is:
\begin{equation}\label{R_L}
R(x,Q^2)=\frac{F_L}{F_T}=\frac{\gamma^2F_2-2xF_1}{2xF_1}=\frac{\gamma^2F_2}{2xF_1}-1.
\end{equation}

\section{Derivative expansion of $F_2$} \label{sec:DE}

The difference between dividing  by the deuteron $F_2^D$  or  by the free isoscalar $F_{2}^N$ structure function to calculate the ratios of structure functions is of only a few percent in the  $x<0.7$ region. However,
the quality of data requires a proper description of $F_2^D$. On the other hand, our local density approach is not appropriate for such a light nucleus (or even for $^4$He).  Therefore, we need another method to calculate $F_2$ in these cases. In terms of the deuteron wave function, $F_2^D$ can be written as
\begin{equation}\label{F2A_Deut}
F^D_2(x,Q^2)=\int \frac{d^3p}{(2\pi)^3}|\Psi_D(\mathbf{p})|^2
\frac{\left(1-\gamma\frac{p_z}{M}\right)}{\gamma^2}
 \left(\gamma'^2+\frac{6x'^2(\mathbf{p}^2-p^2_z)}{Q^2}\right)F_2^N(x',Q^2).
\end{equation}
 Alternatively, a particularly appealing approach because of its simplicity, is the use 
of derivative expansions that provide the structure function per nucleon of a nucleus in terms of the free nucleon one, its derivatives and a few expected values of nuclear observables
\cite{Simo,Akulinichev:1985ij,Frankfurt:1985ui,Kulagin1989,CiofiDegliAtti:1989eg}. We can write
\begin{eqnarray}\label{Ciofi:2007}
F_{2,DEx}^D(x,Q^2) &\simeq&  F_{2}^N(x,Q^2) +
x F_{2}^{N\,'}(x,Q^2)
\frac{ < E > + <T_R>}{M} \nonumber\\
&+& \frac{x^2}{2}
F_{2}^{N\,''}(x,Q^2)\frac{2<T>}{3M}, \label{expNVC1}
\end{eqnarray}
where $<T>$ is the mean nucleon kinetic energy taken as 11.07 MeV, $< E >$ is the nucleon removal energy taken as 2.226 MeV,  $<T_R> \simeq \left< {\bf p}^2\right>/2M$ with $\left< {\bf p}^2\right>= 0.533$ $fm^{-2}$ the average of the square of the nucleon momentum.
To include TMC, one must substitute in Eq.~(\ref{Ciofi:2007}) the free nucleon
structure function and its derivatives by the approximate one given in Eq.~(\ref{f2TMC}).

\begin{figure}
\includegraphics[width=0.75\textwidth]{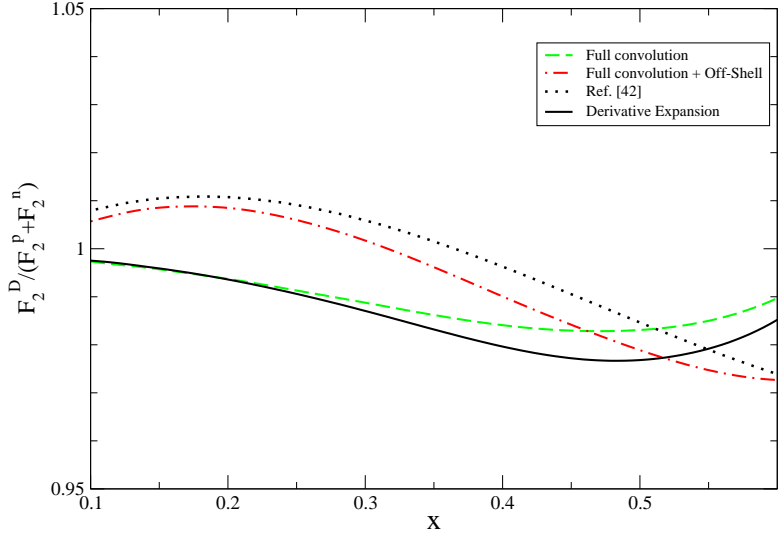}
\caption{$F^D_2/(F_2^p+F_2^n)$ as a function of $x$ at $Q^2$=10 GeV$^2$. Solid line: Derivative expansion.
Dashed line: Eq.~\ref{F2A_Deut}. Dashed dotted:  Eq.~\ref{F2A_Deut} including off-shell effects following the prescription of Ref.~\cite{Petti}. Dotted line:
Ref.~\cite{Kulagin:2010gd}.}\label{fig:f2d}
\end{figure}

The derivative expansions have some intrinsic limitations and it has been shown that they fail to converge to 
the  results obtained by folding with the nuclear spectral functions for $x\gtrsim 0.5$, for the case of medium and heavy nuclei.  A detailed study can be found in Ref.~\cite{Simo}. The convergence is expected to be much better for a loosely bound nucleus such as the deuteron. Indeed,  $F_2^D$ obtained using the Paris wave function~\cite{Paris} and the results of the derivative expansion differ  by less than 0.6 percent up to $x=0.6$ as shown in Fig.~\ref{fig:f2d}. 

In Fig.~\ref{fig:f2d}, we also include for comparison the same ratio from Ref.~\cite{Kulagin:2010gd}, which uses a different set of PDF's.  We have observed that the main difference with that calculation comes from the inclusion of a parametrization of the off-shell effects (see dashed-dotted line) absent in our model.

\section{Results and discussion}\label{sec:RE}
\begin{figure}
\includegraphics[width=0.75\textwidth]{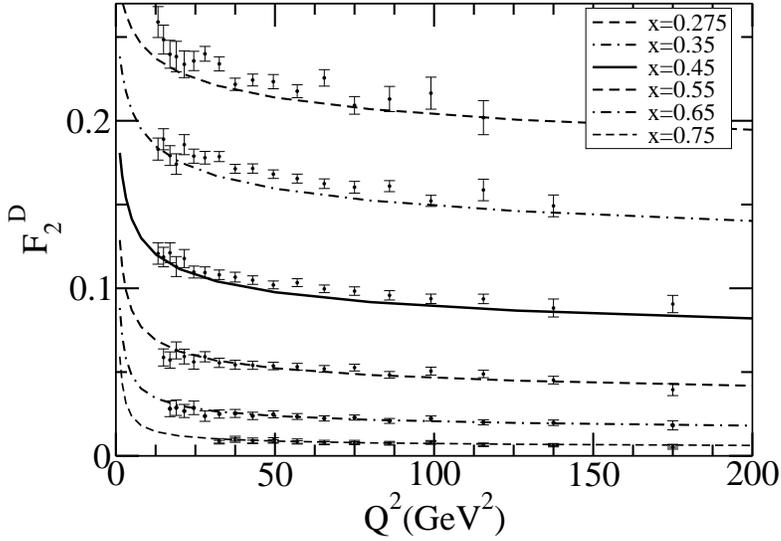}
\caption{Electromagnetic structure function in deuteron $F_{2}^D$ at different $x$ values. The theoretical curves are obtained by using Eq.~(\ref{F2A_Deut}). Experimental data are taken from Ref.~\cite{Benvenuti:BCDMS}.}
\label{fig2}
\end{figure}
Our aim in this paper is to confront the model with the recent JLab results of Ref.~\cite{Seely} that correspond to 
ratios of nuclei with deuteron and more precisely with the slope of the $x$ dependence that is more insensitive to the normalization uncertainties. 
Nonetheless, we will also show some results for the deuteron $F_2^D(x,Q^2)$ structure function as well as the ratio R(x,Q$^2$)=$\frac{2F_{2}^{A}}{AF_2^D}$ in intermediate mass nuclei like $^{40}$Ca and $^{56}$Fe.
\begin{figure}
\includegraphics[width=0.75\textwidth]{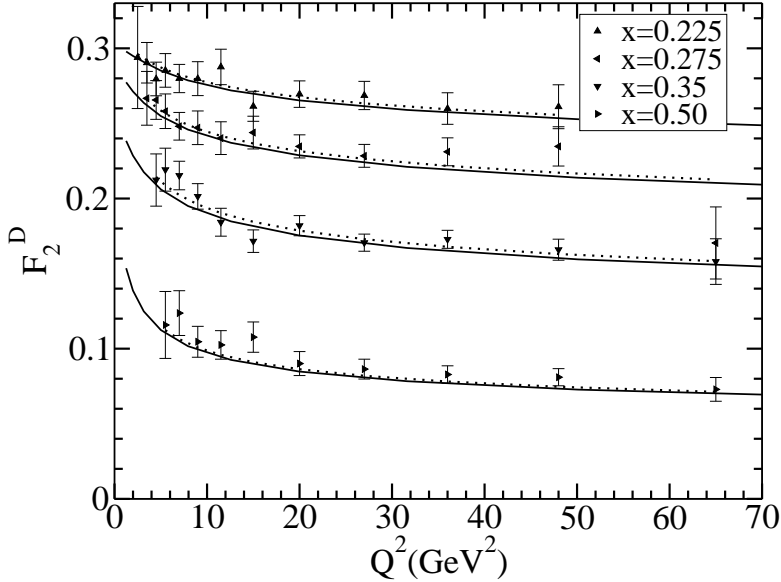}
\caption{Electromagnetic structure function in deuteron $F_2^D$ at different $x$ values. Solid lines are obtained by using Eq.~(\ref{F2A_Deut}) and the dotted lines are the results for the free nucleon case.
 Experimental data are taken from Ref.~\cite{Arneodo:1997}.}\label{fig3}
\end{figure}

In Fig.~\ref{fig2}, we compare the theoretical calculation obtained using Eq. (\ref {F2A_Deut}) with the experimental results of Benvenuti et al.~\cite{Benvenuti:BCDMS}. 
Overall, we find a good agreement in the $x$ region relevant for our study. Although the data correspond to large $Q^2$ values, this gives us confidence in the quality of this approach for the evaluation of the ratios with respect to other nuclei.    
In Fig.~\ref{fig3}, we compare our results  with data obtained with a muon beam on a deuterium target~\cite{Arneodo:1997}. We also show the results  for the free nucleon case. The nuclear corrections are very small for the range of $x$ values analysed in the experiment.

\begin{figure}
\includegraphics[width=0.78\textwidth]{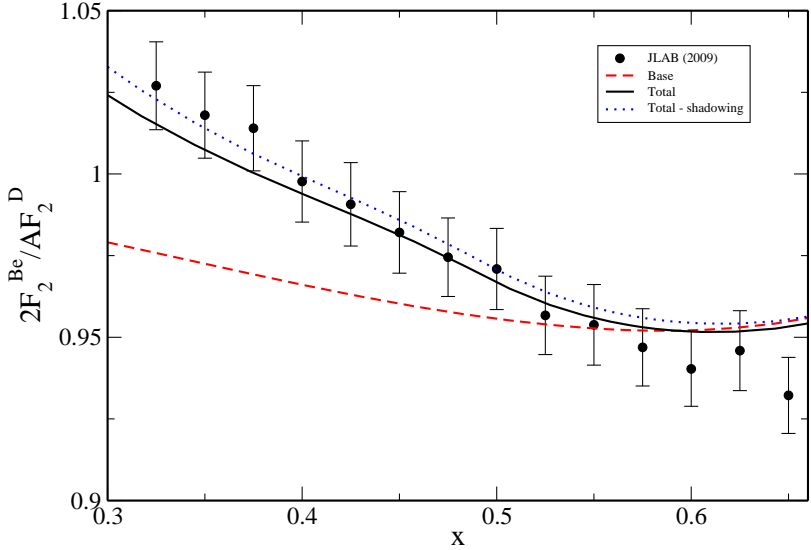}
\caption{Ratio R(x,Q$^2$)=$\frac{2F_{2}^{Be}}{AF_2^D}$. Full model (solid line), without shadowing (dotted line) and
without pion, rho and shadowing contributions (dashed line). For each value of $x$,  $Q^2$ has been calculated using an electron beam of  5.767GeV and scattering angle of $40^0$ corresponding to JLab kinematics. Data are cross section ratios from Ref.~\cite{Seely}.}\label{fig4}
\end{figure}
One of the most interesting results  of the recent JLab data is that for  both Beryllium and Carbon the cross section ratios show a similar slope even when they have a quite different average nuclear density. This  conflicts with some simple fits that describe well the slope for medium and heavy nuclei
as a function of the average nuclear density or with simple A dependences~\cite{Gomez:SLAC}. On the other hand,
the slope of the ratio in the region $0.3<x<0.6$ is particularly well suited to analysis because from the experimental point of view it is quite unaffected by normalization uncertainties. Also theoretically it is relatively simple because shadowing, or Fermi motion are of a little importance over this region of $x$.
\begin{figure}[htb]
\includegraphics[width=0.75\textwidth]{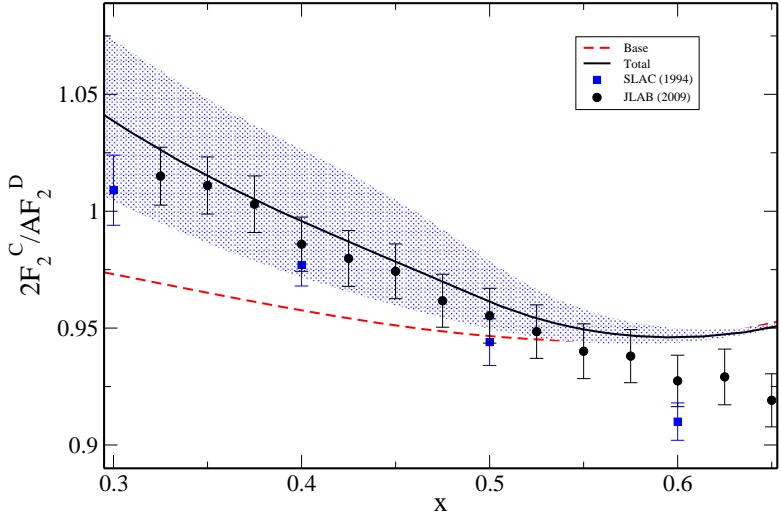}
\caption{Ratio R(x,Q$^2$)=$\frac{2F_{2}^{C}}{AF_2^D}$. Full model: solid line with $\Lambda$, $\Lambda_\rho$ = 1GeV; the band corresponds to $\pm$20\% variation on $\Lambda$ and $\Lambda_\rho$. Full model without  pion, rho and shadowing: dashed line. 
$Q^2$ for calculation and JLab data~\cite{Seely} as described in previous figure. SLAC data~\cite{Gomez:SLAC} correspond to $Q^2=5$ GeV$^2$.} 
\label{fig:RatioCarbon}
\end{figure}

In Fig.~\ref{fig4}, we show the results for Beryllium. The dashed line has been calculated using Eq.~(\ref{F2A_Kulagin}) with TMC and the solid line corresponds to the full model, including the meson cloud contributions, shadowing and TMC. We show explicitly the effect of  shadowing. It reduces the
structure function ratio by around 1\%  at $x\sim0.3$ and even less for higher $x$. We have found that TMC has a really minor effect in the ratio for these $x$ values (less than 1\% at $x\sim0.6$ and even smaller for lower $x$ values). Therefore, the difference between the base curve and the full one comes basically from the $\pi$ and $\rho$ contributions that play an important role.  
The size of the  rho meson correction is about half that of the pion. We find that the full model agrees quite well with data both in slope and the size of the ratio. 

A good agreement with data is also obtained for Carbon as shown in Fig.~\ref{fig:RatioCarbon}. The slope and size of the nuclear effects are similar to the Beryllium case. This could look surprising given the quite different average density as discussed in~\cite{Seely}. This points out to the fact that "average density" could not be the appropriate parameter for the description of the EMC effect in light nuclei. For example, this has been discussed  in Sect. IV of Ref.~\cite{Ciofi degli Atti:2007vx}. Again, a determining factor in the agreement is the mesonic cloud contribution. Given this, some words of caution are needed here. First, the parton distribution functions are poorly known for the mesons and possible off-shell effects have not been included in the calculation.
Second, the results depend on the meson selfenergies in the medium that also contain some uncertainties
 such as the specific form of the spin-isospin interaction, specially for the $\rho$ meson. A full analysis of these uncertainties is out of the scope of this paper. To give an idea of their size, we have shown in this figure
the results for the ratio using the full model with $\Lambda$, $\Lambda_\rho$=1GeV and $\Lambda$, $\Lambda_\rho$=1.2GeV and 0.8GeV. We find that a 20$\%$ variation in the $\Lambda$'s, results in a 2-3$\%$ change in the ratio.

We have also tested the momentum sum rule discussed in section~\ref{sec:meson}. The mesons carry a light-cone momentum fraction of 3 percent for $\Lambda= 1$ GeV. The 2 percent prescribed by the sum rule can be obtained for a cut-off $\Lambda= 0.8$ GeV. This suggests that lower cut-off values should be preferred but one must be careful before reaching such conclusion. For instance, the nucleon momentum fraction is very sensitive to parameters like the expected value of the nucleon kinetic energy that are not very well known and has some uncertainty. We have used the values obtained with our nucleon spectral function. 
The same results are obtained for $^9Be$. Heavier nuclei, such as iron and calcium, have a mesonic  momentum fraction of 5 percent and fulfill the sum rule for $\Lambda= 0.75$ GeV.

In Fig.~\ref{fig:RatioCarbon}, the systematic difference in size between JLab and SLAC data is consistent with the normalization uncertainties quoted in Refs.~\cite{Gomez:SLAC,Seely}. It may be noted, however, that the slope is very similar for both experiments and in good agreement with our results. These normalization differences have been recently discussed in Ref.~\cite{Kulagin:2010gd}.

For both nuclei, our results slightly overestimate data by around 2\%\ at $x$ around 0.6 and more above that. However, that region is much affected by possible off-shell effects~\cite{Kulagin:2010gd}, not included in our approach, and by high momentum components of the nucleons wave function. Therefore, we cannot make any strong statement about this discrepancy apart from the fact that we are reaching one of the limits of validity of our model.

We have also checked that the use of next to next to leading order PDFs, that  considerably lengthens and complicates the calculation, does not appreciably change the results, at the level of precision of the current data and  the size of other theoretical uncertainties.

There are  JLab results even for lighter nuclei like $^3$He and $^4$He. Our local density model is certainly not adequate for these cases that would require a more microscopical approach for the calculation of a proper nucleon spectral function and of the meson cloud contribution.  Also, good data for larger $x$ values are available. They are particularly sensitive to TMC and to high momentum components of the nucleon spectral functions. In order to analyse these data, further work would be required to extend the validity of the theoretical approach describing the nucleon spectral function.

\begin{figure}[h]
\includegraphics[width=0.75\textwidth]{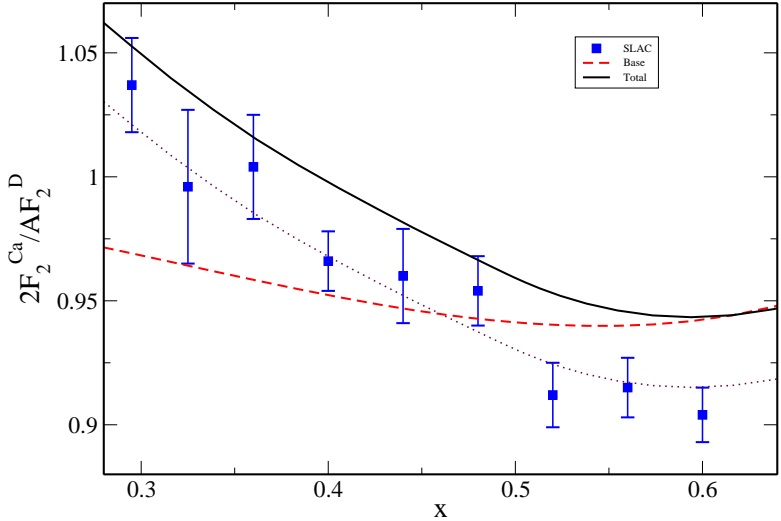}
\caption{Ratio R(x,$Q^2$)=$\frac{2F_{2}^{Ca}}{AF_2^D}$. Full model: solid line. Full model without  pion, rho and shadowing: dashed line.  Dotted curve is the full result scaled by a factor 0.97. Calculations have been done for 
$Q^2=5$ GeV$^2$.
The experimental points are taken from Ref.~\cite{Gomez:SLAC} (averaged $Q^2$).} 
\label{fig:RatioCalcium}
\end{figure}

\begin{figure}[h]
\label{fig7}
\begin{center}
\includegraphics[width=0.75\textwidth]{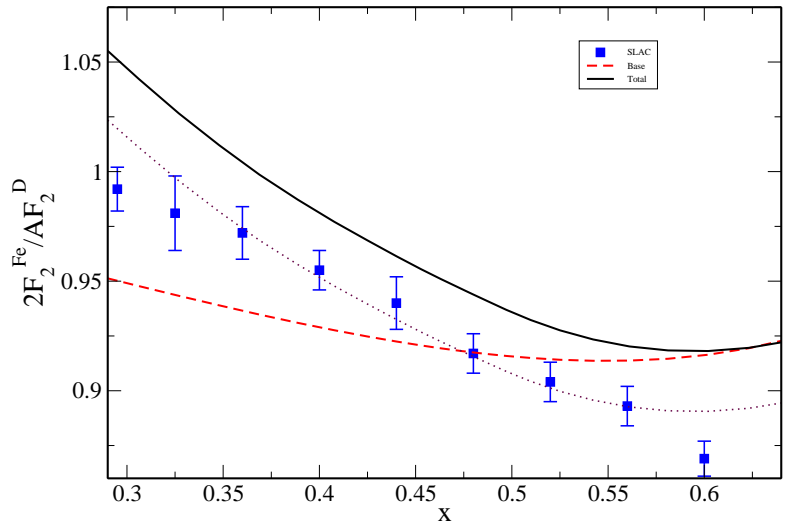}
\caption{ Ratio R(x,Q$^2$)=$\frac{2F_{2}^{Fe}}{AF_2^D}$. Lines have the same meaning as in Fig.~\ref{fig:RatioCalcium}. Calculations have been done for 
$Q^2=5$ GeV$^2$.
Experimental points  from Ref.~\cite{Gomez:SLAC} (averaged $Q^2$).} 
\label{fig:RatioIron}
\end{center}
\end{figure}

As a further test, we have also studied 
the results for the ratio R(x,Q$^2$)=$\frac{2F_{2}^A}{AF_2^D}$ for  intermediate mass nuclei like calcium and iron. The results are shown in Figs. \ref{fig:RatioCalcium} and \ref{fig:RatioIron}. In both cases, we have compared with SLAC results from Ref.~\cite{Gomez:SLAC}, with averaged $Q^2$. No significant $Q^2$ dependence was found in Ref.~\cite{Gomez:SLAC} for their kinematic range. The theoretical curves have been calculated for $Q^2=5$GeV$^2$ and we observe little sensitivity to that value.  In the case of calcium, our results overestimate the data by around  3\%.
This is larger than the normalization uncertainties quoted in Ref.~\cite{Gomez:SLAC}. Nonetheless, scaling our theoretical curve, we observe a good agreement with the slope of the structure function.

The situation is much the same for heavier nuclei, such as iron. The slope is well reproduced and calculation overestimates again data  by around a 3\%. Similar results are obtained for silver and gold. This overestimation seems to be consistent with the results of recent global fits to the  nuclear parton distribution functions (see e.g. Fig. 4 of Ref.~\cite{Schienbein:2009kk}). Their results might point out to some normalization uncertainty in the SLAC results such that the medium and heavy nuclei ratios are too small. However, the recent more microscopical analysis of Ref.~\cite{Kulagin:2010gd} that also finds normalization inconsistencies between light and heavy nuclei favours the interpretation  that the recent JLab data should be rescaled by a global factor of around 0.98 and that the SLAC data are correct.
We should also mention that the two discussed experiments had significantly different lepton energies and  the simple $A$ dependence assumed for the $\sigma_L/\sigma_T$ ratio could have to be revised.

Certainly, these normalization issues should be settled with new and better experiments. From the theoretical point of view, it seems that microscopical models are hardly able to reproduce at the same time the high statistic data from  light nuclei at JLab and medium and heavy nuclei from other collaborations. In any case, this does not affect the main point discussed in this paper, namely the slope produced by the nuclear effects that is well reproduced in our model.

In summary, the electromagnetic nuclear structure function $F_2^A$ has been studied including nucleonic and mesonic degrees of freedom for a $x$ region where shadowing, antishadowing and Fermi motion are not too important. We have started from up to date nucleonic PDF. Nuclear effects like Fermi motion and binding have been incorporated by means of the use of a spectral function obtained for nuclear matter and implemented in nuclei using the local density approximation. A similar approach has been used for the inclusion of the contribution of mesonic clouds. Also shadowing and TMC have been considered. The deuteron structure function has been calculated using a derivative expansion and with the Paris wave function.  The results successfully reproduce recent very precise JLab results for light nuclei at  relatively low $Q^2$  values. Also the slope of previous experiments for heavier nuclei is well reproduced although we fail to agree with them on the absolute size by up to a 3\%, larger than the quoted experimental uncertainty. 
We have found that the mesonic cloud (basically pion) gives an important contribution to the cross section ratios but it still has considerable uncertainties. Even small changes of the pion nuclear selfenergy can produce appreciable changes in  the cross section ratios. 

The success of this local density model for light nuclei is in contrast with the failure of simple models/parametrizations that fit well for the nuclear effects for medium and heavy nuclei as a function of average density or the mass number $A$~\cite{Seely}. The use of an approach that incorporates in an adequate manner the nucleon and meson properties in the nuclei is clearly mandatory for the analysis of the EMC effect in these cases.

\section{Acknowledgments}This work is partly supported by DGICYT contract
number FIS2006-03438. We acknowledge the support of
the European Community-Research Infrastructure Integrating
Activity HadronPhysics2, Grant Agreement n. 227431.
One of us (M. S. A.) wishes to acknowledge the financial support from the University of Valencia and Aligarh Muslim University under the academic exchange program and also to the DST, Government of India for the financial support under the grant SR/S2/HEP-0001/2008. I.R.S. acknowledges Spanish Ministry of Science and Innovation for its support via a FPU grant. We also want to thank Dr. Aji Daniel for fruitful discussions and for his interest in this work.


\begin{thebibliography}{00}

\bibitem{Seely}
  J.~Seely {\it et al.},
  Phys.\ Rev.\ Lett.\  103 (2009) 202301.
 
\bibitem{Aubert:EMC}
  J. J. Aubert et al.,
  Phys. Lett. B 123 (1983) 275. 

\bibitem{Arnold:EMC}
 R. G. Arnold et al.,
 Phys. Rev. Lett. 52 (1984) 727. 

\bibitem{Bodek:EMC}
 A. Bodek et al., 
 Phys. Rev. Lett. 51 (1983) 534. 

\bibitem{Gomez:SLAC} 
  J. Gomez et al., 
  Phys. Rev. D 49 (1994) 4348.

\bibitem{Arneodo:1994}
M. Arneodo, Phys. Rep. 240 (1994) 301.

\bibitem{Geesaman:1995}
  D.~F.~Geesaman, K.~Saito and A.~W.~Thomas,
  Ann. Rev. Nucl. Part. Sci. 45 (1995) 337.

\bibitem{Armesto:2006ph}
  N.~Armesto,
  J. Phys. G 32 (2006) R367.
  
\bibitem{Miller:EMCTh}
  G.A. Miller, 
  Eur. Phys. J. A 31 (2007) 578. 

\bibitem{Hirai:NPDF}  M.~Hirai, S.~Kumano and T.~H.~Nagai,
  Phys.\ Rev.\  C { 76} (2007) 065207.

\bibitem{Eskola:NPDF} K.J. Eskola, H. Paukkunen and C.A. Salgado,
JHEP 0904 (2009) 065.

\bibitem{Schienbein:NPDF} I. Schienbein et al.,
Phys.\  Rev.\  D 77 (2008) 054013.

\bibitem{Schienbein:2009kk}
  I.~Schienbein, J.~Y.~Yu, K.~Kovarik, C.~Keppel, J.~G.~Morfin, F.~Olness and J.~F.~Owens,
  Phys.\ Rev.\  D  80 (2009) 094004.


\bibitem{Marco} E. Marco, E. Oset and P. Fernandez de Cordoba,
   Nucl. Phys. A 611 (1996) 484. 


\bibitem{FernandezdeCordoba:1991wf}
  P.~Fernandez de Cordoba and E.~Oset,
  Phys. Rev. C 46 (1992) 1697.


\bibitem{GarciaRecio:1994cn}
  C.~Garcia-Recio, J.~Nieves and E.~Oset,
  Phys.\ Rev.\  C { 51} (1995) 237.


\bibitem{Schienbein:2007gr}
  I.~Schienbein {\it et al.},
  J.\ Phys.\ G 35 (2008) 053101.

\bibitem{Petti} S. A. Kulagin and R. Petti, 
  Nucl. Phys. A 765 (2006) 126.

\bibitem{Petti2} S. A. Kulagin and R. Petti, 
   Phys. Rev. D 76 (2007) 094033.    

\bibitem{MSTW} A. D. Martin, W.J. Stirling, R. S. Thorne and G. Watt,
    hep-ph:0901.0002 http://durpdg.dur.ac.uk/hepdata/mrs.html. 

\bibitem{Vermaseren} J. A. M. Vermaseren et al., 
  Nucl. Phys. B 724 (2005) 3.
  
\bibitem{Neerven} W. L. van Neerven and A. Vogt,
   Nucl. Phys. B 568 (2000) 263; ibid 588 (2000) 345.

\bibitem{Gluck:1991ey}
  M.~Gluck, E.~Reya and A.~Vogt,
  Z.\ Phys.\  C 53 (1992) 651.

\bibitem{Gluck} M. Gluck, E. Reya and I. Schienbein,
  Eur. Phys. J. C 10 (1999) 313.
  
\bibitem{Callan} C. G. Callan, Jr. and D. J. Gross,
   Phys. Rev. Lett. 22 (1969) 156.

\bibitem{Moch:2004xu}
  S.~Moch, J.~A.~M.~Vermaseren and A.~Vogt,
  Phys.\ Lett.\  B 606 (2005) 123.

\bibitem{De Jager:1987qc}
  H.~De Vries, C.~W.~De Jager and C.~De Vries,
  Atom.\ Data Nucl.\ Data Tabl.\  { 36} (1987) 495.


\bibitem{Simo} I. Ruiz Simo and M.J. Vicente Vacas,
  J. Phys. G 36 (2009) 015104.   

  
\bibitem{Ankowski:2007uy}A.~M.~Ankowski and J.~T.~Sobczyk,
  Phys. Rev. C 77 (2008) 044311.   
      
\bibitem{Sajjad} M. Sajjad Athar, S.K. Singh and M.J. Vicente Vacas,
	Phys. Lett. B 668 (2008) 133. 


\bibitem{Oset:1981ih}
  E.~Oset, H.~Toki and W.~Weise,
  Phys.\ Rept.\  { 83} (1982) 281.


\bibitem{Carrasco:1989vq}
  R.~C.~Carrasco and E.~Oset,
  Nucl.\ Phys.\  A { 536} (1992) 445.

\bibitem{Nieves:1991ye}
  J.~Nieves, E.~Oset and C.~Garcia-Recio,
  Nucl.\ Phys.\  A { 554} (1993) 554.


\bibitem{Nieves:1993ev}
  J.~Nieves, E.~Oset and C.~Garcia-Recio,
  Nucl.\ Phys.\  A { 554} (1993) 509.

\bibitem{Gil:1997bm}
  A.~Gil, J.~Nieves and E.~Oset,
  Nucl.\ Phys.\  A { 627} (1997) 543.

\bibitem{Ciofi degli Atti:2007vx}
  C.~Ciofi degli Atti, L.~L.~Frankfurt, L.~P.~Kaptari and M.~I.~Strikman,
  Phys. Rev.  C 76 (2007) 055206.

\bibitem{Akulinichev:1985ij}
  S.~V.~Akulinichev, S.~A.~Kulagin and G.~M.~Vagradov,
  Phys.\ Lett.\  B { 158} (1985) 485.


\bibitem{Frankfurt:1985ui}
  L.~L.~Frankfurt and M.~I.~Strikman,
  Phys.\ Lett.\  B { 183}, 254 (1987).

\bibitem{Kulagin1989}
S.~A.~Kulagin,
Nucl. Phys. A 500 (1989) 653.

\bibitem{CiofiDegliAtti:1989eg}
  C.~Ciofi Degli Atti and S.~Liuti,
  Phys.\ Lett.\  B { 225} (1989) 215.


\bibitem{Paris}
  M.~Lacombe, B.~Loiseau, R.~Vinh Mau, J.~Cote, P.~Pires and R.~de Tourreil,
  Phys.\ Lett.\  B { 101} (1981) 139.

\bibitem{Kulagin:2010gd}
  S.~A.~Kulagin and R.~Petti,
Phys. Rev. C 82 (2010) 054614.

\bibitem{Benvenuti:BCDMS}
  A. C. Benvenuti et al., 
  Phys. Lett. B { 237} (1990) 592. 

\bibitem{Arneodo:1997} M. Arneodo et al.
  Nucl. Phys. B { 483} (1997) 3.  


\end{thebibliography}
\end{document}